\documentclass[%
 reprint,
showpacs,
 amsmath,amssymb,
 aps,
]{revtex4-1}
\usepackage{graphicx}
\usepackage{dcolumn}
\usepackage{bm}

\usepackage{natbib}
\usepackage[lofdepth,lotdepth]{subfig}
\usepackage{xcolor}
\usepackage{placeins}
\newcommand{\ti}{\textit}
\newcommand{\tb}{\textbf}
\newcommand{\mr}{\mathrm}

\begin{document}

\preprint{APS/123-QED}
\title{Loss of classicality in alternating spin-$\frac{1}{2}$/spin-$1$ chain, in the presence of 
next-neighbor couplings and Dzyaloshinskii-Moriya interactions}

\author{Abhiroop Lahiri}
 \altaffiliation[Also at ]{Chemistry and Physics of Materials Unit, JNCASR, Bangalore India}
\author{Swapan K Pati}%
 \email{pati@jncasr.ac.in}
\affiliation{%
 Theoretical Sciences Unit, Jawaharlal Nehru Centre for Advanced Scientific Research, Bangalore, India }%

\date{\today}

\begin{abstract}
We have considered and alternating spin-$\frac{1}{2}$/spin-$1$ chain
with nearest-neighbor ($J_1$), next-nearest neighbor ($J_2$) 
antiferromagnetic Heisenberg interactions along
with z-component of the Dzyaloshinskii-Moriya(DM)
($D_z$) interaction.
The Hamiltonian has been studied using (a) Linear Spin-Wave Theory(LSWT) and (b) Density Matrix Renormalization Group (DMRG).
The system had been reported earlier as a classical ferrimagnet only when nearest neighbor exchange interactions are present. Both the 
antiferromagnetic next-nearest neighbor interactions and DM 
interactions introduce strong quantum fluctuations and due to which
all the signatures of ferrimagnetism vanishes. We find that the
nonzero $J_2$ introduces strong quantum fluctuations in each of
the spin sites due to which the z-components of both spin-1 and 
spin-1/2 sites average out to be zero. The ground state 
becomes a singlet. The presence of $J_1$ along with $D_z$ introduces a short range order but develops long range order along the XY plane. $J_1$ along with $J_2$ induces competing phases with structure 
factor showing sharp and wide peaks, at two
different angles reflecting the spin spiral structure locally 
as well as in the underlying lattice. Interestingly, we find that 
the $D^z$ term removes the local spin spiral structure in z-direction, while developing a spiral order in the XY plane.
 
\end{abstract}
\pacs{02.70.-c, 67.80.Jd}
\maketitle


\section{\label{sec:level1}INTRODUCTION}

Low dimensional quantum spin systems are of considerable interest 
as they exhibit a wide range of exotic physical phenomena\cite{low_d_1, low_d_2}. 
Due to strong 
quantum fluctuations, in most of the quantum low dimensional systems,
the long range order gets destroyed even at absolute zero temperature.
Many such properties have theoretically and computationally been 
predicted \cite{low_d_5, low_d_6, low_d_7, low_d_8} and most of those have 
already been realized experimentally\cite{expt1, expt2, expt3, expt4, 
expt5, expt6}. The most 
popular model to explain 
these phenomena is the Heisenberg model along with various other terms 
such as anisotropy \cite{aniso1, aniso2, aniso3, aniso4}, spin-orbit coupling and 
Dzyaloshinskii-Moriya \cite{dm1, dm1A} interactions. Most of these terms
break the SU(2) symmetry and lead to exotic broken
symmetry ground states. 

  The low-lying excitations in these systems vary dramatically depending 
on the site spins (whether they are integer or half-odd-integer) and 
nature of the 
superexchange interactions. The nearest neighbor integral spin or 
nearest and next-neighbor coupled half-odd-integer spins show a finite
gap in the excitation spectrum and in effect a short range two-spin correlation
functions in the ground state. Interestingly, the next-neighbor 
coupling induces frustration
in the system and such an infinite degeneracy of the classical ground state
of the spin system gets lifted when quantum fluctuations are introduced.
The gapless spin system, the spin gap due to resonating valence bond or 
spontaneous dimerization due to frustration, single magnon state, 
multi-magnon states, spin-glass and spin-ice ground states in a large classes
of magnetic systems have already been realized computationally and 
experimentally. There have been studies on the low-energy and low 
temperature properties of alternating spin chains with nearest neighbor 
Heisenberg interactions. Such ferrimagnetic systems have
been shown to display a rich low energy spectrum with both antiferromagnetic (AFM)
ground state and ferromagnetic excitations. Due to underlying non compensating
site spins with finite magnetization, these low dimensional systems show
long range magnetic order with finite magnetization even with finite temperature.Since quantum fluctuations of the Heisenberg model cannot destroy the classical ferromagnetic order, these alternating spin-chains can we well expalined in the limit of Linear Spin Wave Theory (LSWT).

    In this work, we are interested in the effect of 
Dzyaloshinskii-Moriya (DM) interactions on the low energy spectrum
of Heisenberg chain systems consisting of spin$\frac{1}{2}$ and spin-$1$ at alternating sites (Fig \ref{alt_spin}
with competing exchange interactions. 
We have carried out detailed studies on the ground state properties of these systems using perturbative 
LSWT and non-perturbative Density Matrix Renormalization
Group (DMRG) methods. We have compared the spin-density,
two-spin correlation functions and the structure factor between
the perturbative and non-perturbative methods and with change in 
the magnitude of alternating site spins in the low-dimensional systems.

   In next section, we have carried out detailed analysis of the
low-energy spectrum of an alternating spin-$\frac{1}{2}$/spin-$1$ chain ($S_1, S_2$) with
nearest neighbor AFM interactions and
the z-component of the DM interactions using LSWT. In the subsequent section, DMRG calculations have been performed
on the alternating spin systems  (and at
times with next-nearest neighbor frustrated term). The ground state energy,
spin density and two-point equal time correlation functions and various
other order parameters have been calculated to characterize the ground state.
We conclude with summary of all the results in the last section of the manuscript.
 
\section{RESULTS WITH NEAREST-NEIGHBOR INTERACTIONS}
\subsection{\label{sec:level2}Linear Spin Wave Theory analysis}

The Heisenberg Hamiltonian for an alternating chain of spins of $S_1$ 
and $S_2$ with the z-component of the Dzyaloshinskii-Moriya (DM) interactions can be 
written as 
 \begin{multline}\label{eq:hamil}
\mathcal{H}  = J\sum_{ i }\left\{\bm{S}_{1,i}.\bm{S}_{2,i} + \bm{S}_{2,i}.\bm{S}_{1,i+1}\right\} \\
   + {D}^z \sum_{ i }\biggl\{S^+_{1,i}S^-_{2,i}
   -S^-_{1,i}S^+_{2,i}
   + S^+_{2,i}S^-_{1,i+1} \\
   + S^+_{2,i}S^-_{1,i+1} \biggr\}
 \end{multline}

A detailed calculation has been carried out, which can be found in the Supplementary Information
 
We apply the Holstein-Primakoff transformations\\
\begin{equation}\label{eq:hp1}
\begin{aligned}
S^z_{1,i} = S_1 - a^{\dagger}_ia_i, \\
S^+_{1,i} = \sqrt{2S_1-a^{\dagger}_ia_i}a_i,\\
S^-_{1,i} = a^{\dagger}_i \sqrt{2S_1-a^{\dagger}_ia_i},
\end{aligned}
\end{equation}
for the spin $S_1$ sites and \\
\begin{equation}\label{eq:hp2}
\begin{aligned}
S^z_{2,i} = -S_2 + b^{\dagger}_ib_i, \\
S^+_{2,i} = b^{\dagger}_i \sqrt{2S_2-b^{\dagger}_ib_i},\\
S^-_{2,i} = \sqrt{2S_2-b^{\dagger}_ib_i}b_i
\end{aligned}
\end{equation}\\
\vspace{1.0cm}
for the spin $S_2$ sites. Substituting the above in 
Eq.\ref{eq:hamil}, we obtain the Hamiltonian in in the representation of the Holstein-Primakoff bosons. A subsequent Fourier transformation leads to
\begin{align}
   \mathcal{H}_{1k}
   &=2\sqrt{S_1S_2}\sum_{k}\mr{cos}(\frac{k}{2})\left(J^{(+)}a_kb_{-k} + J^{(-)}a^{\dagger}_kb^{\dagger}_{-k}\right) \notag\\
&+ J\sum_{k}\{S_2\left(a^{\dagger}_ka_k +a^{\dagger}_{-k}a_{-k}\right) 
 + S_1\left(b^{\dagger}_kb_k + + b^{\dagger}_{-k}b_{-k}\right) \}
\end{align}
In the basis
$\bm{A}_k = \{a^{\dagger}_k, b^{\dagger}_{k}, a_{-k}, b_{-k} \}$, this can be written in matrix form
\begin{equation}
    \mathcal{H}_{1k} = \sum_k \bm{A}^{\dagger}_k \bm{H}_k \bm{A}_k
\end{equation}
where \\
\vspace{2.5cm}
\begin{widetext}
\begin{align}
\bm{H}_k 
&=
\begin{pmatrix}
 JS_2 & 0 & 0 & J^-\sqrt{S_1S_2}\mr{cos}(\frac{k}{2}) \\
 0 &  JS_1 & J^-\sqrt{S_1S_2}\mr{cos}(\frac{k}{2}) & 0 \\
 0 & J^+\sqrt{S_1S_2}\mr{cos}(\frac{k}{2}) & JS_2 & 0 \\
 J^+\sqrt{S_1S_2}\mr{cos}(\frac{k}{2}) & 0 & 0 & JS_1
  \end{pmatrix}
\end{align}
\end{widetext}
Bogoliubov transformation, followed by diagonalization 
(see Supplementary Information) gives 2 doubly degenerate modes modes :
\begin{equation}
\begin{aligned}
{\omega_{1k}} = \frac{1}{2}\left\{\left(S_1 - S_2\right) \right\} - \Omega_k\\
{\omega_{2k}} = \frac{1}{2}\biggl[\left(S_1 - S_2\right) \biggr] + \Omega_k\\
\end{aligned}
\end{equation}\\
where
\begin{equation}\label{eq:discri-1}
    \Omega_k =\frac{1}{2}\sqrt{-4\left(J^2 + D^2\right)S_1S_2
    \mathrm{cos}^2(k/2) + J^2(S_1+S_2)^2}
\end{equation}\\

\begin{figure}[h]
\centering
\subfloat[][]{
\includegraphics[width=0.24\textwidth]{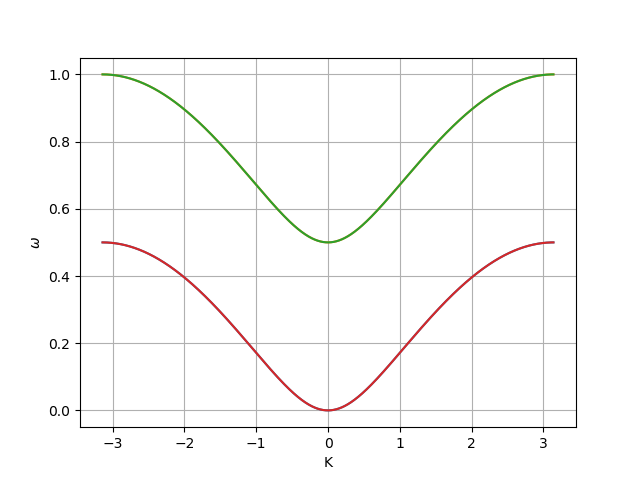}
\label{fig:subfig1}}
\hspace{-2.0em}
\subfloat[][]{
\includegraphics[width=0.24\textwidth]{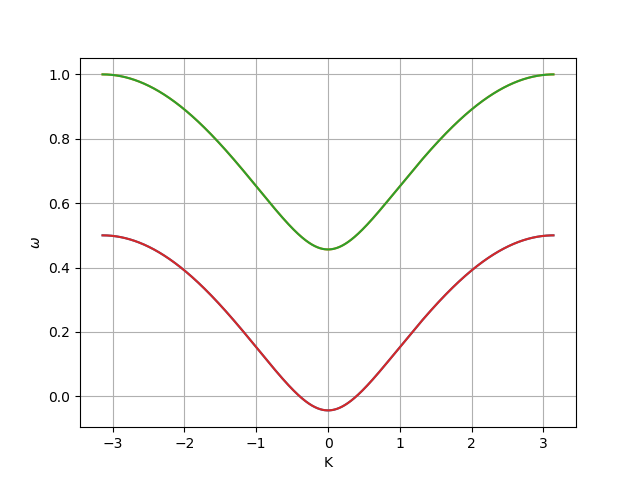}
\label{fig:subfig2}}\\
\subfloat[][]{
\includegraphics[width=0.24\textwidth]{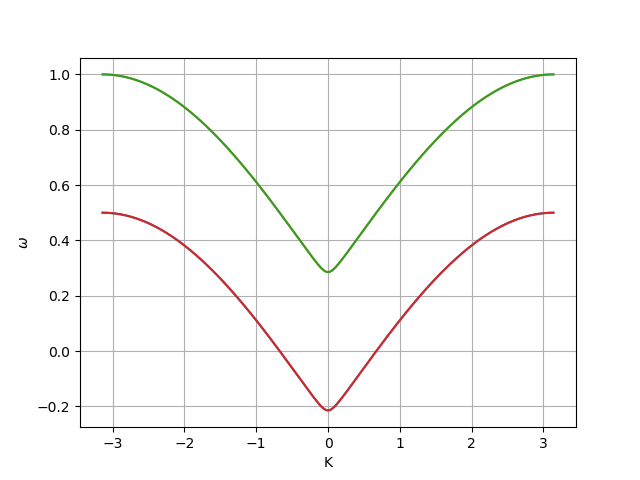}
\label{fig:subfig3}}
\hspace{-2.0em}
\subfloat[][]{
\includegraphics[width=0.24\textwidth]{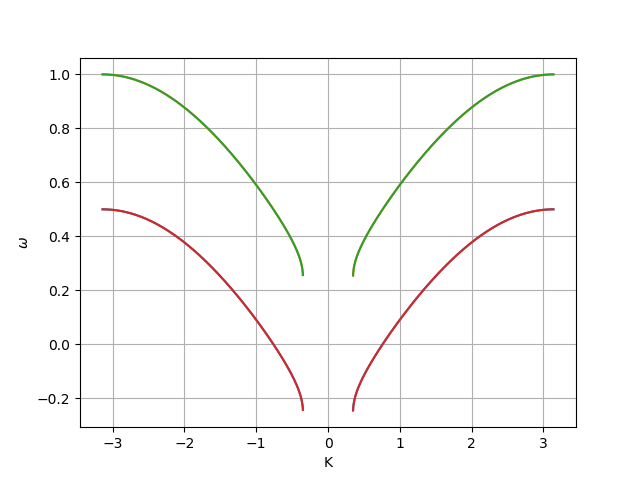}
\label{fig:subfig4}}
\caption{Energy dispersion plots for $D/J$ = (a) 0.0, (b)0.2, (c) 0.35 and 
(d) 0.4 for a spin-$\frac{1}{2}$/spin-$1$ alternating chain}
\label{fig:magdis1}
\end{figure}

Two low-energy spin wave dispersion curves are shown in Fig. \ref{fig:magdis1}, corresponding to the two sublattices with two different
bosonic modes. Here, we have dropped the superscript of the DM interaction, i.e., $D^z = D$. In the case of $D=0$, the lower dispersion mode (colored red) is the 
gapless mode, due mainly to the antiferromagnetic interactions, while the higher energy
dispersion mode is gapped and has features of ferromagnetic interactions. This same relation was obtained by Pati et. al \cite{pati}. As $D$ is increased, 
within the linear spin-wave theory, the dispersion $\omega_{1k}$ becomes negative 
near $k=0$. This implies that there is an 
instability with respect to static spin-wave 
formalism\cite{nega-sw}. This instability can be 
removed by applying an external magnetic field as there will be an extra tunable 
coefficient $B$ of $a_k^{\dagger}a_k$ due to the field, which will shift the value 
of $\omega_{1k}$ back to zero. We will see below how the non-perturbative method accounts
for such instability when the quantum fluctuations is properly accounted for.

 For $D>0.3535$, the discriminant, $\Omega_{k}^2$ (\ref{eq:discri-1}) becomes negative. 
Thus, both the $\omega$'s, namely, $\omega_{1k}$ and $\omega_{2k}$ become complex. 
Hence, the spin wave dispersion curves become unrealistic and are thus no longer valid. 
The sublattice magnetizations can be calculated as the 
expectation values of $S^z_{1i}$ and $S^z_{2i}$.
\begin{figure}[h!]
\centering

\includegraphics[width=0.24\textwidth]{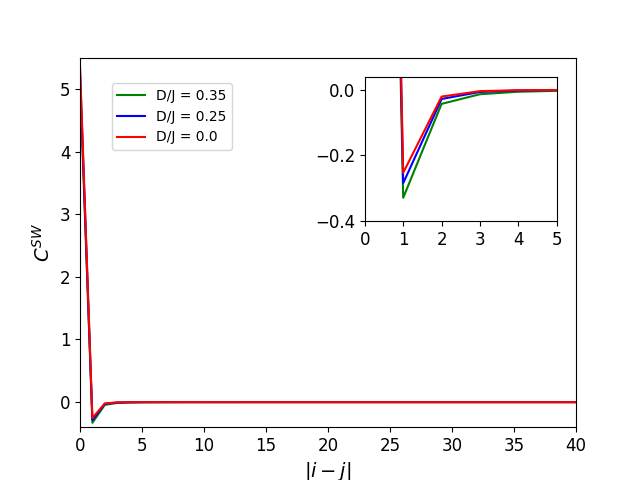}
\caption{Spin correlation function plots from linear spin wave theory for (a) nearest-neighbour and (b) next-nearest-neighbour interactions in a spin-$\frac{1}{2}$/spin-$1$ alternating chain}
\label{fig:corr_swn}
\end{figure}

 \begin{align}\label{eq:magsw1}
    M_a 
    &= S_1 - \langle {a^{\dagger}}_{i=1}a_{i=1}\rangle
    = S_1 - \frac{1}{N_{uc}}\sum_{k,k'}\langle a^{\dagger}_{k}a_{k'}\rangle
\end{align}
and 
 \begin{align}\label{eq:magsw2}
    M_b 
    &= \langle {b^{\dagger}}_{i=1}b_{i=1}\rangle - S_2
    = \frac{1}{N_{uc}}\sum_{k,k'}\langle b^{\dagger}_{k}b_{k'}\rangle - S_2
\end{align}
We can calculate (Suppl.)
\begin{align}\label{eq:magsw3}
  \frac{1}{N_{uc}}\sum_{k,k'}\langle b^{\dagger}_{k}b_{k'}\rangle
&= \frac{1}{N_{uc}}\sum_{k,k'}\langle a^{\dagger}_{k}a_{k'}\rangle \notag\\
= \left(S_1 + \frac{1}{2}\right) &+ \frac{1}{2}\int_{-\pi}^{\pi}\frac{J(S_1+S_2) + \Omega_k}{4(J^2+D^2)S_1S_2\mathrm{cos}^2(k/2)}dk
\end{align}

As can be seen from linear spin-wave results, for each 
spin-$\frac{1}{2}$/spin-$1$ dimer, the total magnetization is $(S_1-S_2) = \frac{1}{2}$ 
since the fluctuations in each of the site spins are exactly opposite 
and thus gets cancelled out to give finite magnetization value for every dimer. Thus, the 
system behaves as a classical ferrimagnet with alignment of finite magnetizations 
of dimers in a lattice. However, this result is valid only 
for $D=0$, as evident from non perturbative DMRG calculations, discussed 
in the next subsection. For $D\ne 0$, the non-perturbative quantum 
fluctuations make the spin of each site as well as the dimer to be zero.

The $S^z$-$S^z$ correlation function is given by
\begin{equation}
    C^z\left(\vert i-j\vert\right) = 
    \langle S^z_iS^z_j \rangle - \langle S^z_i \rangle\langle S^z_j\rangle
\end{equation}

These correlation functions comprise of 3 types, $\langle S^z_{1i}S^z_{1j} \rangle$, 
$\langle S^z_{2i}S^z_{2j} \rangle$ and $\langle S^z_{1i}S^z_{2j} \rangle$. 
These are plotted in Fig. X 
Since the system behaves as a ferrimagnet with finite magnetization for 
every dimer, the correlation functions should be calculated after subtracting 
the product of the averages of each $S^z$ values, given by

\begin{equation}
    \langle S^z_{1i}S^z_{1j} \rangle - \langle S^z_{1i}\rangle\langle S^z_{1j} 
\rangle=  \frac{1}{2\pi}\int_{-\pi}^{\pi}
    \left\{1 + \frac{\Omega_k\mathrm{cos}(k\vert i-j\vert }{J(S_1+S_2)}\right\} dk
\end{equation}
\noindent for the first type of correlations.

The plots of the correlation function with distance between the spin sites 
are shown in Fig \ref{fig:corr_swn}. This clearly shows very short range order upto 
only a few sites, which is consistent with the earlier work \cite{pati}
for $D=0$. But for $D \ne 0$, in the LSWT regime, the correlation length is also very 
small and there is not much significant difference between the correlation function 
for different values of $D$ until the function becomes complex at $D \simeq 0.36$. 
Thus LSWT fails to explain this case, as it assumes that there is primarily 
antiferromagnetic order with fluctuations, even in presence of $D$. 
This is contrary to all the DMRG results, especially with nonzero $D$, presented later. 
\begin{figure}[h!]
\centering
\subfloat[][]{
\includegraphics[width=0.24\textwidth]{corr_sw.png}
\label{fig:corr_swn}}
\hspace{-2.0em}
\subfloat[][]{
\includegraphics[width=0.24\textwidth]{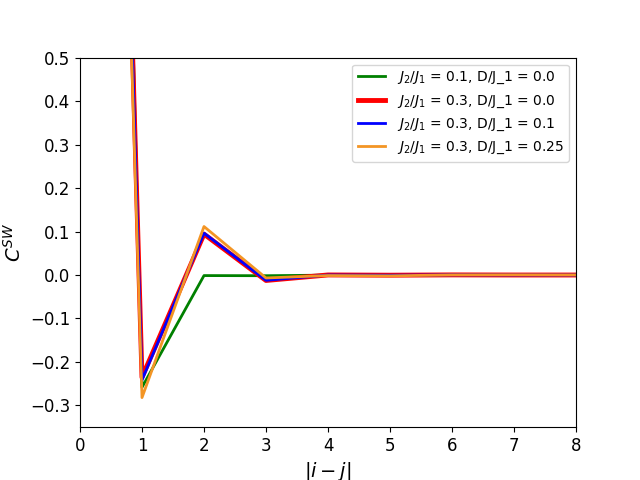}
\label{fig:corr_swnn}}\\
\caption{Spin correlation function plots from linear spin wave theory for (a) nearest-neighbour and (b) next-nearest-neighbour interactions in a spin-$\frac{1}{2}$/spin-$1$ alternating chain}
\end{figure}

\vspace{-1em}
\subsection{\label{sec:citeref}DMRG Results}

\begin{figure}[h!]
\centering
\subfloat[][]{
\includegraphics[width=0.24\textwidth, height=0.149\textheight]{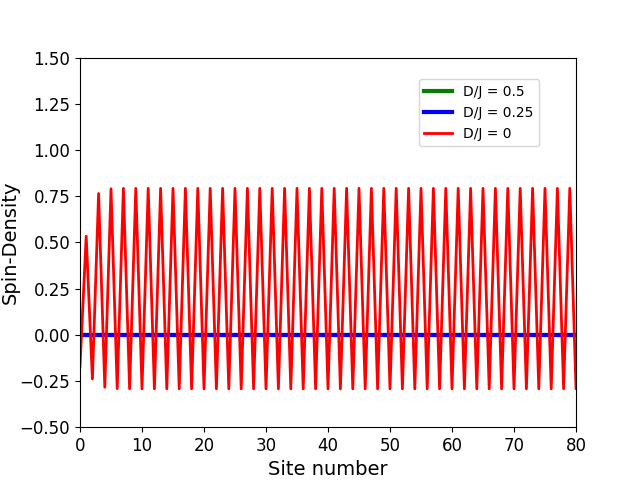}
\label{fig:spnden}}\hspace{-2em}
\subfloat[][]{
\includegraphics[width=0.24\textwidth]{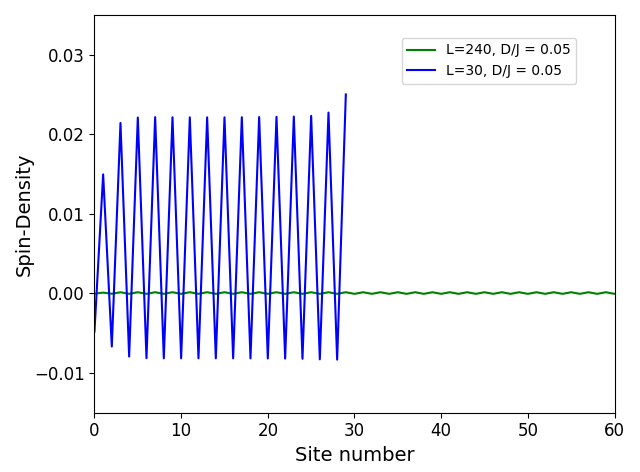}\label{fig:spnden005}}\\
\vspace{-1.0em}
\subfloat[][]{
\includegraphics[width=0.24\textwidth]{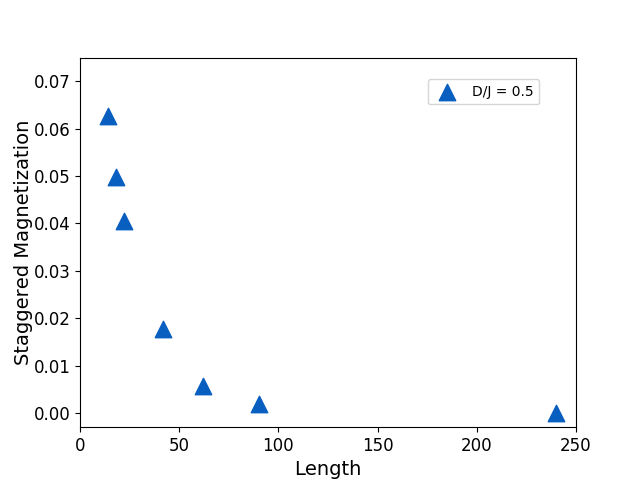}
\label{fig:den_vs_l}}
\caption{Plots of (a)Spin density vs site for a 240 site spin-$\frac{1}{2}$/spin-$1$ alternating chain for $D^z/J= 0.0, 0.25, 0.5$
(b) Spin density for $D^z/J = 0.05$ for 30 and 240 sites, and, 
(c) Staggered magnetization for various lengths for $D^z/J=0.5$,
all with NN interactions}
\label{spin_densities}

\end{figure}

Finite size DMRG calculations have been performed in 
the Matrix Product States(MPS) formalism for a  
chain of 120 and 240 sites with nearest neighbor (NN) as well 
as next-nearest neighbor (NNN) magnetic exchange interactions 
and the z-component of the Dzyaloshinskii-Moriya interactions. In this paper we present the results for 240 sites. 
The cut off for bond dimension 
of the MPS has been kept to be $500$ and finite size sweeps upto 200 have been used to obtain the converged ground state.
Without DM interactions, the spin density (shown in Fig.\ref{fig:spnden} ) of 
each site is less than the classical value, but the
difference between spin density for each dimer is 1/2 as expected, 
which is also confirmed 
from our LSWT calculations and previous work\cite{pati}. This means 
that each site has quantum fluctuations, but each 
spin-$\frac{1}{2}$-spin-$1$ dimer has classical magnetization value. \\
Hence the $S^z$-$S^z$ correlation function $C^z\left(\vert i-j\vert\right)$
sharply falls to zero after a length of two sites, as the product of the 
averages have finite values. This is the characteristic of a magnetic chain
with long range order, which in this case is due to the formation of a 
ferrimagnetic chain with finite dimer magnetization in the lattice.

\begin{figure}[h!]
\centering
\subfloat[][]{
\includegraphics[width=0.24\textwidth]{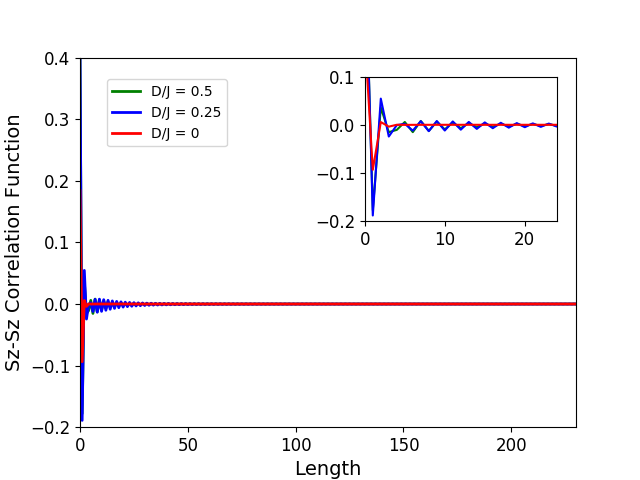}
\label{fig:szcor240}}
\hspace{-2.0em}
\subfloat[][]{
\includegraphics[width=0.24\textwidth]{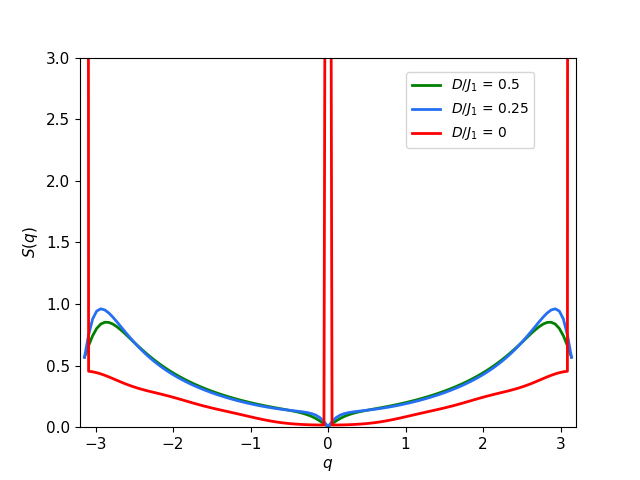}
\label{fig:szstrtuc240}}
\caption{Plots of (a) correlation function $C^z\left(\vert i-j\vert\right)$, 
(b) structure factor $S(q)$ vs $q$, (c) spin density vs site and (d) $C^{(+)}\left(\vert i-j\vert\right)$
$D/J$ = 0.0, 0.25 and 0.5 for a spin-$\frac{1}{2}$/spin-$1$ 
alternating chain of 240 sites with NN interactions}
\end{figure}

However, the moment we turn on the nearest-neighbor 
DM interactions, it introduces strong quantum fluctuations 
in each of the spin sites, making the dimer move away from
classical magnetic state. Although, the quantum fluctuation was
present in each site spin when $D=0$, the dimer did not
have any, because oppositely oriented z-component of the 
site spins had exactly opposite quantum fluctuations, thus cancelling each other. However, with $D \ne 0.0$, the situation is very
different; it not only introduces strong quantum fluctuation in each 
of the site spins, their z-components vanishes thereby the
dimer z-component also vanishes. This makes each spin density at 
individual sites zero even though
there are different site spins in every alternate sites. Interestingly,
the quantum fluctuations are more for larger magnitude spins and LSWT fails to explain any of these even with large magnitude 
of DM interactions. This suggests that although in a ferrimagnet, the 
interactions are antiferromagnetic, any non-local 
interactions can destroy the site magnetization of different magnitude 
magnetic ions as well as the dimeric magnetiation of classically 
non-compensating spin dimers. This is also manifested in the correlation function $C^z\left(\vert i-j\vert\right)$ which decays at a slower rate\ref{fig:szcor240}, thus introducing 
quasi long range order as the product of the averages are zero. \\
The classical ferrimagnetic state of the alternating spin $\frac{1}{2}$/spin-$1$ chain is unstable even for a small value of $D^z/J$.
 As can be seen, for the nonzero $D/J$, the spin density at every site vanishes. To verify the quantum fluctuation at every site spin, we have considered small and large sized lattice and have calculated spin density for a small $D/J$ value. It is clear from Fig. \ref{fig:spnden005} that, for $D/J = 0.05$, the magnetization fluctuates antiferromagnetically, however, the scale of fluctuation is quite small (of the order of $~10^{-2}$). However, as we increase the lattice length, the site magnetization vanishes. Thus, in the thermodynamic limit, when the system size goes to infinity, the magnetization will go to zero eventually, even for small DM interactions. In Fig. \ref{fig:den_vs_l} the staggered magnetization $\frac{1}{L}\displaystyle\sum_{i=1}^L \vert (-1)^i \langle S^z_i \rangle \vert$ vs chain length has been plotted for a large value $D/J=0.5$, which also clearly shows the nonlinear decrease in magnitude of magnetization with lattice length.

The destruction of classical ferrimagnetic ordering in the chain is also 
reflected in the structure factor, $S(q)$, for $D=0$, shown in Fig. \ref{fig:szstrtuc240}. The 
preferred direction of ordering of any two spin-$\frac{1}{2}$ 
or spin-$1$ at alternate sites is parallel. This is manifested 
in the sharp peaks of $S(q)$ at $q=0$ and at $q=\pi$. 
For $D\ne 0$, the peak at $q=0$ or at $q=\pi$ vanishes, as 
there is no preferred ordering. Nevertheless, there are 
two broad peaks at $q<\pi$ and $q>-\pi$, referring to 
some canted ordering angle. In fact, the spiral ordering angle
varies with the magnitude of $D$.\\
\begin{figure}[h!]
\centering
\subfloat[][]{
\includegraphics[width=0.24\textwidth]{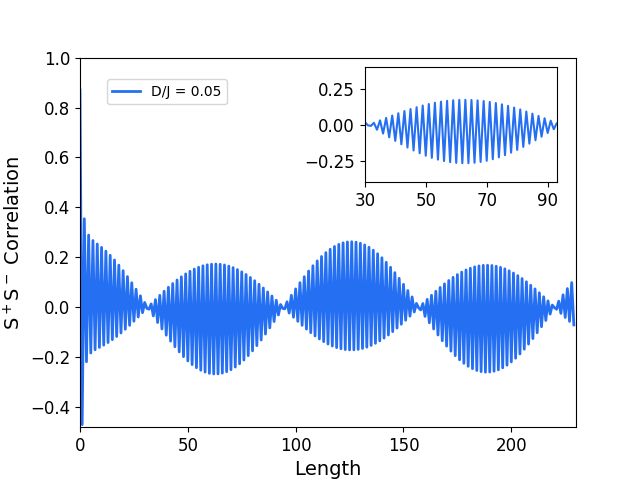}
\label{fig:spsm005}}
\hspace{-2.0em}
\subfloat[][]{
\includegraphics[width=0.24\textwidth]{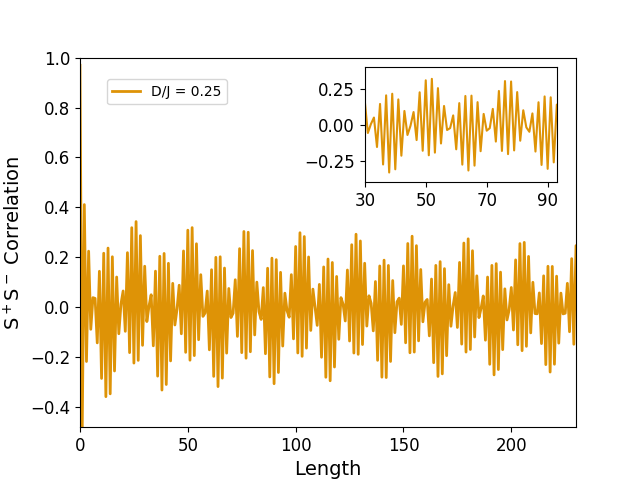}
\label{fig:spsm025}}\\
\vspace{-1em}
\subfloat[][]{
\includegraphics[width=0.24\textwidth]{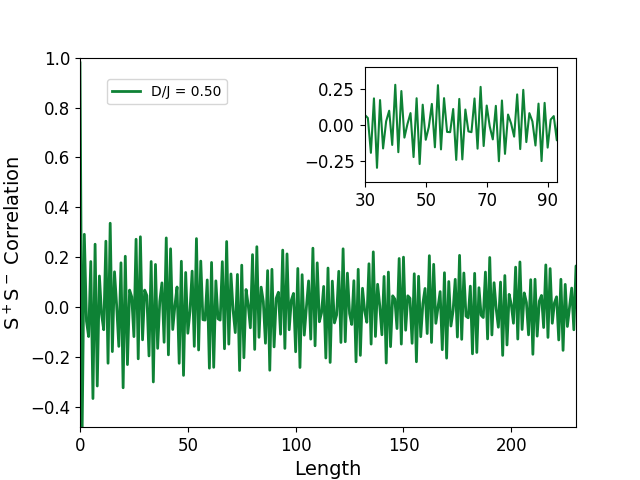}
\label{fig:spsm050}}
\caption{Plots of correlation function  $C^{(+)}\left(\vert i-j\vert\right)$ for
$D/J$ = 0.05 (a), 0.25 (b) and 0.5 (c) for a spin-$\frac{1}{2}$/spin-$1$ 
alternating chain of 240 sites with NN interactions  }
\label{fig:spsm_all}
\end{figure}
The $S^+$-$S^-$ correlation given by
\begin{equation}
C^{(+)}\left(\vert i-j\vert\right) = \langle S^+_i S^-_j \rangle - \langle S^+_i \rangle\langle S^-_j\rangle
\end{equation}
\noindent starts building up with finite values for nonzero $D$, which on the other hand 
was decaying for $D=0$. This correlation function has fluctuations, which continue
for longer distances.
Interestingly, within
each of the spiral order, the fluctuation is maximum in the middle and 
there appears to be a periodicity, which varies with the variation of the DM strength (Fig.\ref{fig:spsm_all}).\\

\section{RESULTS WITH NEXT-NEAREST-NEIGHBOR INTERACTIONS}
\subsection{Spin-Wave Theory Analysis}
The Hamiltonian for next-nearest neighbor (NNN) 
exchange interaction in k-space is given by

\begin{equation*}
    \mathcal{H}_{2k} = \mathcal{H}_{1k}+\mathcal{H}_{nn,k}
\end{equation*}

\noindent where
\begin{equation}
 \mathcal{H}_{nn,k} = J_2\sum_{\delta,k}\left( e^{2ik\delta} -1  
\right)\left(S_1a^{\dagger}_ka_k + S_2b^{\dagger}_kb_k\right)
\end{equation}

Bogoliubov transformation, followed by diagonalization 
(see Supplementary Information) gives 2 modes as before (the other 
two are the same):

\begin{equation}
\begin{aligned}
{\bm{\omega_{1k}}} = \frac{1}{4}\biggl[\left(S_1 - S_2\right)\left\{ J_1 
+ J_2\left(1-2\mr{cos}(2k)\right) \right\} \biggr] + \Omega_k\\
{\bm{\omega_{2k}}} = \frac{1}{4}\biggl[\left(S_1 - S_2\right)\left\{ J_1 
+ J_2\left(1-2\mr{cos}(2k)\right) \right\} \biggr] - \Omega_k\\
\end{aligned}
\end{equation}
\noindent where
\begin{widetext}
\begin{equation*}
    \Omega_k = \sqrt{-4\left(J_1^2 + D^2\right)S_1S_2
    \mathrm{cos}^2(k) + 
    \left\{J_1 - 2J_2\left(\frac{1}{2}-\mr{cos}(2k)\right)\right\}^2(S_1+S_2)^2 }
\end{equation*}
\end{widetext}
A similar expression without the DM interaction had been derived by Mohakud et. al \cite{pati2}.
The energy dispersion modes for different parameters are plotted in Fig \ref{fig:corr_swnn}.\\
On introducing the next-neighbor AFM coupling, one introduces spontaneous frustration 
in a one-dimensional antiferromagnetic lattice .
This leads to the dispersion relation becoming flattened 
and as the $J_2$ is increased further, the spin group velocity reduces. On further increase of $J_2$, the AFM dispersion mode flattens more, until it 
becomes negative at $J_2=0.25$ and complex.
The magnetization in this case (calculated similarly as in 
Eqs. \ref{eq:magsw1}, \ref{eq:magsw2}, \ref{eq:magsw3}) reduces from its 
classical value, but for the dimer, it remains the 
same, $\left(S_1-S_2\right)$, which is classical. 

The correlation function vs. length is plotted in Fig \ref{fig:corr_swnn}. 
Clearly, the LSWT predicts short range order for both non-zero $J_2$ and nonzero 
$D$ values, contrary to DMRG results.
\begin{figure}[h!]
\centering
\subfloat[][]{
\includegraphics[width=0.24\textwidth]{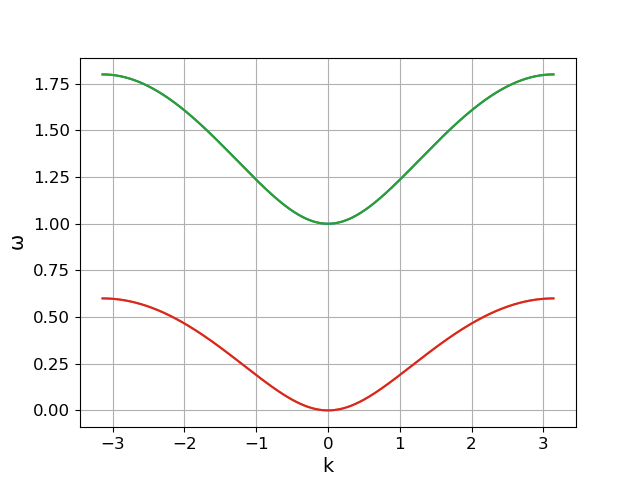}
\label{fig:subfig1}}
\hspace{-2.0em}
\subfloat[][]{
\includegraphics[width=0.24\textwidth]{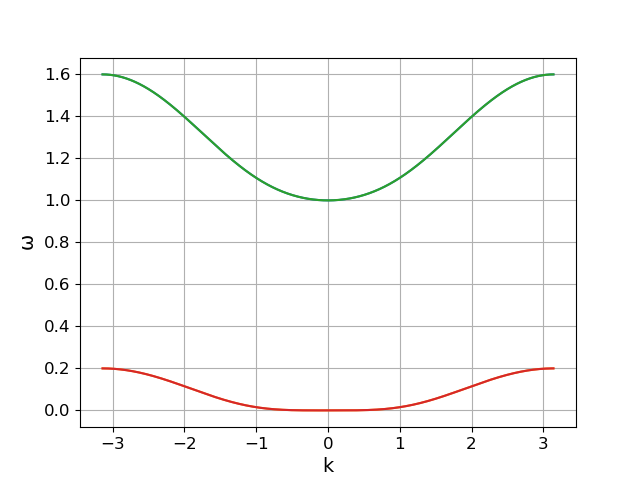}
\label{fig:subfig2}}\\
\caption{Energy dispersion plots for spin-$\frac{1}{2}$/spin-$1$ alternating 
chain with next-neighbor frustration for $D/J_1$ = (a) 0.0, (b)0.2, (c) 0.35 and (d) 0.4 }
\label{fig:magdis3}
\end{figure}
\FloatBarrier

\subsection{DMRG Results}
\begin{figure}[h!]
\centering
\subfloat[][]{
\includegraphics[width=0.24\textwidth]{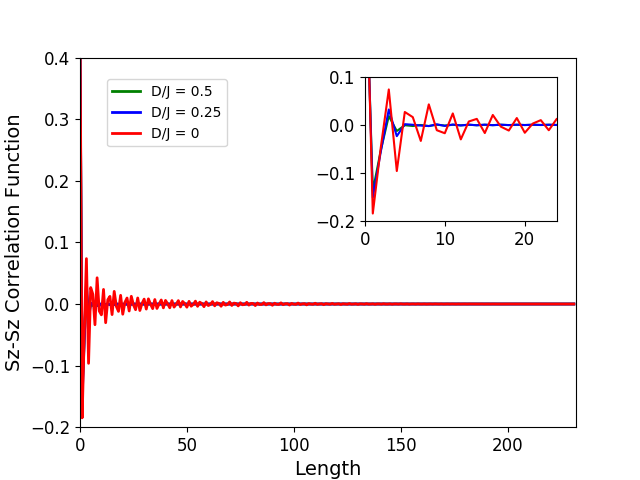}
\label{fig:szcor240_n}}
\hspace{-2.0em}
\subfloat[][]{
\includegraphics[width=0.24\textwidth]{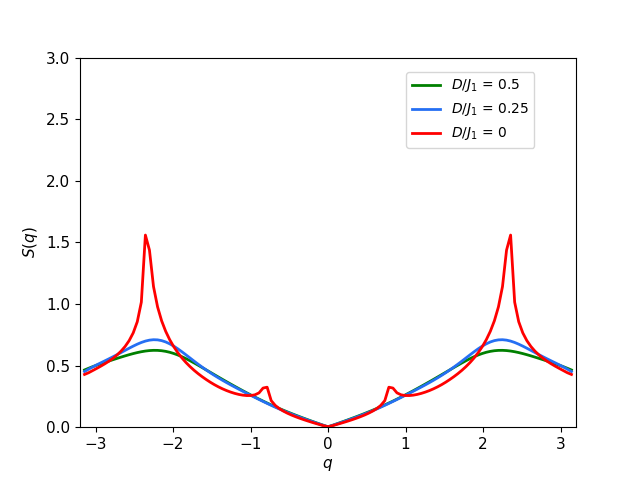}
\label{fig:szstruc240_n}}\\
\vspace{-1em}
\subfloat[][]{
\includegraphics[width=0.48\textwidth]{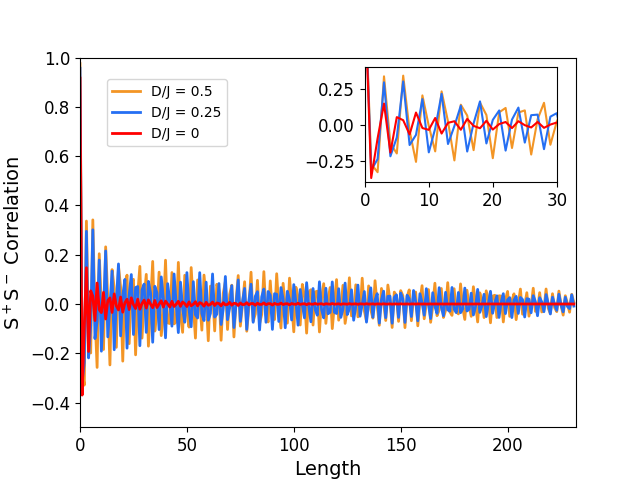}
\label{fig:spsmcor240_n}}
\caption{Plots of (a) correlation function $C^z\left(\vert i-j\vert\right)$, 
(b) structure factor $S(q)$ vs $q$ and (c) $C^{(+)}\left(\vert i-j\vert\right)$ 
for $J_2/J_1 = 0.4$ and $D/J_1$ = 0.0, 0.25 and 0.5 for a spin-$\frac{1}{2}$/spin-$1$ 
alternating chain of 240 sites with NNN interactions}
\end{figure}

The DMRG calculations were performed for $J_2/J_1 = 0.4$. 
It was shown, as earlier\cite{pati2}, that 
dimer no longer behaves as classical magnetic dimer with finite magnetization value. 
The magnetization average at each site 
becomes zero again, thus making the magnetization value in the dimer 
to be zero. For $D=0$, the frustration propagates and quasi long range order is introduced, as evident 
from the $C^z\left(\vert i-j\vert\right)$ correlation function plotted in Fig. \ref{fig:szcor240_n}. As $D$ 
is turned on, unlike 
that of only nearest neighbor interactions, the quasi long range 
order is destroyed. Hence, in 
both the cases, the DM interaction term changes the spin vectors 
through quantum fluctuations and thereby the two point correlation function
between them. Thus it destroys completely or to some extent the 
quasi long range order or short range order set by the Heisenberg 
interaction terms. Here also, with small values of DM interactions
the spontaneous changes in quantum fluctuations in spin density and spin spin
correlation functions occur.\\

The correlation function $C^{(+)}\left(\vert i-j\vert\right)$ builds up more with increase in $D$, as shown in Fig. \ref{fig:spsmcor240_n}.
Interestingly, for $D/J_1 = 0$, and with nonzero $J_2$,
both $C^z\left(\vert i-j\vert\right)$ and 
$C^{(+)}\left(\vert i-j\vert\right)$ show spiral ordering.
However, the $J_2$ introduces frustration and due to this, the
local as well as the global spin ordering changes. 

As can be seen from the correlation functions, the 
antiferromagnetic short range correlations between two 
consecutive spin
sites (from a given site) remain positive or negative, suggesting
local frustrated ferrimagnetic domains.
This is manifested in the structure factor, $S(q)$, where
we find two sharp peaks, each at $\pi/2>q>0$ and $0<q<\pi$ 
(shown in Fig \ref{fig:szstruc240_n}). For $D\ne 0$, this local order vanishes, again giving two broad peaks at 
$-\pi>q>0$ and $0<q<\pi$. The structure factor reveals the 
manifestation of the competing nature of the frustrated 
interactions and z-component of the DM interactions. Note that,
the DM interactions arise due to the local non-centrosymmetry in the 
spin systems\cite{dm2,dm3,dm4}. \\
Thus, from
Fig. \ref{fig:szcor240_n}, \ref{fig:szstruc240_n} and \ref{fig:spsmcor240_n}
is clear that since we have considered DM interactions along the XY plane (with 
D component along z direction), it preserves the local spiral character for spin 
components along the plane, but kills the z-component local spiral order. Furthermore,
investigation reveals that the local spiral angle is different with the magnitude of 
$D$ values, since the "same sign correlations" appear at different distances 
for varying $D$ values. Thus, the main point is that although $J_2$ 
introduces frustration and thereby degeneracy and local order, the DM 
interactions along the in-plane direction preserves the local order, while 
along the z-direction kills the local order. 
\section{CONCLUSIONS}
We have investigated the ground state and low energy properties of an alternating 
spin-$\frac{1}{2}$/spin-$1$ chain in the presence of DM interactions 
and next-neighbor frustrations. Without DM and next-neighbor antiferromagnetic 
interactions, both Linear Spin Wave 
Theory and non-perturbative DMRG results predict the ground state to be a 
classical ferrimagnetic state with ith total spin $N(S_1-S_2)$. When DM 
interactions are present, however small it may be, LSWT again predicts a 
ground state with total spin $N(S_1-S_2)$, contrary to DMRG results, which 
show that the ground state to have total spin zero. The reason behind this 
is that the DM interactions introduce strong quantum fluctuations at 
each site, thus making the spin at each site zero, and hence the total 
spin zero. This effect could be captured by cubic or quartic orders of 
Spin Wave Theory \cite{sw_dm}, which will be addressed after further work.
In presence of next-neighbor frustration, the system again goes away
from classical limit, and the average of z-component of each site spin becomes
zero and two point equal time correlations functions show quasi-long range 
order. Both next-nearest neighbor frustration and DM interactions destroy this spiral order, and also any kind of short range order along the z-axis.

\clearpage
\end{document}


\maketitle

\section{Spin wave theory analysis}
The Heisenberg Hamiltonian for an alternating chain of spins $s_1$ and $S_2$ with Dzyaloshinskii-Moriya (DM) interactions can be written as 

 \begin{align}
  \mathcal{H} 
  &= J\sum_{\langle ij \rangle}\bm{S}_i.\bm{S}_j + \bm{D}.\bm{S}_i\times\bm{S}_j \\
   &= J\sum_{\langle ij \rangle}S^z_i S^z_j + \frac{J}{2}\left(S^+_i S^-_j + S^-_i S^+_j\right) + \frac{iD^z}{2}\left(S^+_i S^-_j - S^-_i S^-_j\right)
 \end{align}
 for $\bm{D} = \{D^z,0,0\}$\\
Here the A sublattice comprise of spin-$\frac{1}{2}$ sites and the B sublattice comprise of spin-$1$. We apply the Holstein-Primakoff transformations\\
\begin{minipage}{0.45\textwidth}
\begin{center}
    \textbf{\underline{Sublattice A}}
\end{center}
\begin{equation}
\begin{aligned}
S^z_{1,i} = S_1 - a^{\dagger}_ia_i, \\
S^+_{1,i} = \sqrt{2S_1-a^{\dagger}_ia_i}a_i,\\
S^-_{1,i} = a^{\dagger}_i \sqrt{2S_1-a^{\dagger}_ia_i},
\end{aligned}
\end{equation}
\end{minipage}
\hfill
\begin{minipage}{0.45\textwidth}
\vspace{2em}
\begin{center}
    \textbf{\underline{Sublattice B}}
\end{center}
\begin{tabular}{|p{\textwidth}}
\begin{equation}
\begin{aligned}
S^z_{2,i} = -S_2 + b^{\dagger}_ib_i, \\
S^+_{2,i} = b^{\dagger}_i \sqrt{2S_2-b^{\dagger}_ib_i},\\
S^-_{2,i} = \sqrt{2S_2-b^{\dagger}_ib_i}b_i
\end{aligned}
\end{equation}
\end{tabular}
\end{minipage}
Substituting the above equations we get

\begin{align}
\mathcal{H}
&=J\sum_{\langle ij \rangle}\biggl[\left(S_1 -  a^{\dagger}_ia_i\right)\left(b^{\dagger}_jb_j-S_2\right)
+ \frac{1}{2}\biggl(a_i\sqrt{2S_1-a^{\dagger}_ia_i}b_j\sqrt{2S_2-b^{\dagger}_jb_j}
\notag\\
&\qquad \qquad \qquad \qquad + a^{\dagger}_i\sqrt{2S_1-a^{\dagger}_ia_i}b^{\dagger}_j\sqrt{2S_2-b^{\dagger}_jb_j} \biggr)\biggr]
\notag\\
&+ \frac{iD^z}{2} \sum_{\langle ij \rangle}
\biggl(a_i\sqrt{2S_1-a^{\dagger}_ia_i}b_j\sqrt{2S_2-b^{\dagger}_jb_j}
- a^{\dagger}_i\sqrt{2S_1-a^{\dagger}_ia_i}b^{\dagger}_j\sqrt{2S_2-b^{\dagger}_jb_j} \biggr)
\notag\\\\
&\approx 2NJS_1S_2 + J\sum_{\langle ij \rangle}\biggl[S_1b^{\dagger}_jb_j + S_2a^{\dagger}_ia_i + \sqrt{S_1S_2}\left(a_ib_j + a^{\dagger}_ib^{\dagger}_j\right) \biggr]
\notag\\
&\qquad \qquad \qquad
+ iD_z\sqrt{S_1S_2}\sum_{\langle ij \rangle} \left(a_ib_j - a^{\dagger}_ib^{\dagger}_j\right)
\notag\\
&= \mathcal{H}_0 + \mathcal{H}_1
\end{align}\\
After the applying Fourier transform, we get $\mathcal{H} = \mathcal{H}_{cl} + \mathcal{H}_{1k}$ where
\begin{multline}
\mathcal{H}_{1k} = \sqrt{S_1S_2}\sum_{\delta,k}\left(J^{(+)}e^{-ik\delta}a_kb_{-k} + J^{(-)}e^{ik\delta}a^{\dagger}_kb^{\dagger}_{-k}\right)
+ 2J\sum_{k}\left(S_2a^{\dagger}_ka_k + S_1b^{\dagger}_kb_k\right)
\end{multline}
where $\delta$ is the sum of all nearest neighbours.\\
$J^{(+)} = J + iD^z$ and $J^{(-)} = J - iD^z$ . \\
For a 1D chain with nearest neighbours placed equidistantly,$\sum_{\delta,k}e^{ik\delta} = \sum_{\delta,k}e^{-ik\delta} = 2\sum_{k}\mr{cos}(\frac{k}{2})$\\
So, we get,
\begin{align}\label{k_ham}
\mathcal{H}_{1k}
&= 2\sqrt{S_1S_2}\sum_{k}\mr{cos}(\frac{k}{2})\left(J^{(+)}a_kb_{-k} + J^{(-)}a^{\dagger}_kb^{\dagger}_{-k}\right)
+ 2J\sum_{k}\left(S_2a^{\dagger}_ka_k + S_1b^{\dagger}_kb_k\right) 
\notag\\
&=2\sqrt{S_1S_2}\sum_{k}\mr{cos}(\frac{k}{2})\left(J^{(+)}a_kb_{-k} + J^{(-)}a^{\dagger}_kb^{\dagger}_{-k}\right)
+ J\sum_{k}\{S_2\left(a^{\dagger}_ka_k +a^{\dagger}_{-k}a_{-k}\right) \notag\\
&\qquad + S_1\left(b^{\dagger}_kb_k + + b^{\dagger}_{-k}b_{-k}\right) \}
\notag\\
&=\begin{pmatrix}
 JS_2 & 0 & 0 & J^-\sqrt{S_1S_2}\mr{cos}(\frac{k}{2}) \\
 0 &  JS_1 & J^-\sqrt{S_1S_2}\mr{cos}(\frac{k}{2}) & 0 \\
 0 & J^+\sqrt{S_1S_2}\mr{cos}(\frac{k}{2}) & JS_2 & 0 \\
 J^+\sqrt{S_1S_2}\mr{cos}(\frac{k}{2}) & 0 & 0 & JS_1
  \end{pmatrix}
\end{align}
The above matrix is written in the basis
$\bm{A}_k = \{a^{\dagger}_k, b^{\dagger}_{k}, a_{-k}, b_{-k} \}$

\subsection*{Bogoliubov Transformation and Diagonalization}
The Bogoliubov transformation can be written in a generalized form as:
\begin{equation}
    \bm{A}_k = \bm{V}_k .\bm{\Tilde{A}}_k
\end{equation}
where $\bm{\Tilde{A}}_k$ is the array of 4 Bogoliubov operators and $\bm{V}_k$ is the coefficient matrix.
The diagonal form can be obtained by the similarity transformation
\begin{equation}\label{eq:diag}
 \bm{V}^{\dagger}_k \mathcal{H}_k \bm{V}_k = \bm{\Omega}_k
\end{equation}
Also, the commutation relations are preserved after transformation into the Bogoliubov basis. So,
\begin{equation}\label{eq:commutation}
    \bm{g} = \bm{V}_k\  \bm{g}\  \bm{V}^{\dagger}_k
\end{equation}
where $\bm{g}$ is the commutator matrix, written as 
\begin{equation}
    \bm{g} = \bm{A}_k \bm{A}_k^{\dagger} - \left[\left(\bm{A}_k^{\dagger}\right)^T
    \bm{A}_k^T\right]^T
    = \begin{pmatrix}
      \mathbb{1}_2 & \ \\
       \ & -\mathbb{1}_2
      \end{pmatrix}
\end{equation}

From Eq. (\ref{eq:diag}) and (\ref{eq:commutation}), we get
\begin{equation}
    \left(\bm{g}\mathcal{H}_k\right).\bm{V}_k = \bm{V}_k. \left(\bm{g}\bm{\Omega}_k\right)
\end{equation}
This is and eigenvalue equation. The columns of similarity matrix $\bm{V}_k$ are the eigenvectors of $\left(\bm{g}\mathcal{H}_k\right)$. \\
Diagonalizing $\left(\bm{g}\mathcal{H}_k\right)$, and then multiplying by $\bm{g}^{-1}(=\bm{g})$ we get the 4 modes,
\begin{equation}
    \omega_{1k} = \omega_{4k}= \frac{1}{2}\left\{J\left(S_1 - S_2\right) - \sqrt{-4\left(J^2 + D^2\right)S_1S_2
    \mathrm{cos}^2(k/2) + J^2(S_1+S_2)^2}\right\}
\end{equation}
and 
\begin{equation}
    \omega_{2k} = \omega_{3k} = \frac{1}{2}\left\{J\left(S_1 - S_2\right) + \sqrt{-4\left(J^2 + D^2\right)S_1S_2
    \mathrm{cos}^2(k/2) + J^2(S_1+S_2)^2}\right\}
\end{equation}

\subsection{Next-Nearest Neighbour Interactions}
\begin{align}
\mathcal{H}
&=J_1\sum_{\langle ij \rangle}\biggl[\left(S_1 -  a^{\dagger}_ia_i\right)\left(b^{\dagger}_jb_j-S_2\right)
+ \frac{1}{2}\biggl(a_i\sqrt{2S_1-a^{\dagger}_ia_i}b_j\sqrt{2S_2-b^{\dagger}_jb_j}
\notag\\
&\qquad \qquad \qquad \qquad + a^{\dagger}_i\sqrt{2S_1-a^{\dagger}_ia_i}b^{\dagger}_j\sqrt{2S_2-b^{\dagger}_jb_j} \biggr)\biggr]
\notag\\ 
&+J_2\sum_{\langle\langle ij \rangle\rangle}\biggl[\left(S_1 -  a^{\dagger}_ia_i\right)\left(S_1-a^{\dagger}_ja_j\right)
+ \frac{1}{2}\biggl(\sqrt{2S_1-a^{\dagger}_ia_i}a_ia^{\dagger}_j\sqrt{2S_1-a^{\dagger}_ja_j}
\notag\\
&\qquad \qquad \qquad \qquad + a^{\dagger}_i\sqrt{2S_1-a^{\dagger}_ia_i}\sqrt{2S_1-a^{\dagger}_ja_j}a_j \biggr)\biggr]
\notag\\
&+J_2\sum_{\langle\langle ij \rangle\rangle}\biggl[\left(b^{\dagger}_ib_i -S_2\right)\left(b^{\dagger}_jb_j - S_2\right)
+ \frac{1}{2}\biggl(b^{\dagger}_i\sqrt{2S_1-b^{\dagger}_ib_i}\sqrt{2S_1-b^{\dagger}_jb_j}b_j
\notag\\
&\qquad \qquad \qquad \qquad + \sqrt{2S_2-a^{\dagger}_ia_i}b_ib^{\dagger}_j\sqrt{2S_2-b^{\dagger}_jb_j} \biggr)\biggr]
\notag\\
&+ \frac{iD^z}{2} \sum_{\langle ij \rangle}
\biggl(a_i\sqrt{2S_1-a^{\dagger}_ia_i}b_j\sqrt{2S_2-b^{\dagger}_jb_j}
- a^{\dagger}_i\sqrt{2S_1-a^{\dagger}_ia_i}b^{\dagger}_j\sqrt{2S_2-b^{\dagger}_jb_j} \biggr)
\notag\\
&\approx -2NJ_1S_1S_2 + J_1\sum_{\langle ij \rangle}\biggl[S_1b^{\dagger}_jb_j + S_2a^{\dagger}_ia_i + \sqrt{S_1S_2}\left(a_ib_j + a^{\dagger}_ib^{\dagger}_j\right) \biggr]
\notag\\
&\qquad \qquad \qquad
+ iD_z\sqrt{S_1S_2}\sum_{\langle ij \rangle} \left(a_ib_j - a^{\dagger}_ib^{\dagger}_j\right)
\notag\\
&+NJ_2S_1^2 + J_2\sum_{\langle\langle ij \rangle\rangle}S_1\biggl[-a^{\dagger}_ia_i - a^{\dagger}_ja_j + a_ia^{\dagger}_j + a^{\dagger}_ia_j  \biggr]
\notag\\
&+NJ_2S_2^2 + J_2\sum_{\langle\langle ij \rangle\rangle}S_2\biggl[-b^{\dagger}_ib_i - b^{\dagger}_jb_j + b_ib^{\dagger}_j + b^{\dagger}_ib_j  \biggr]
\end{align}\\
After the Fourier transformation, we get
\begin{multline}
\mathcal{H}_{1k} = \sqrt{S_1S_2}\sum_{\delta,k}\left(J^{(+)}e^{-ik\delta}a_kb_{-k} + J^{(-)}e^{ik\delta}a^{\dagger}_kb^{\dagger}_{-k}\right)\\
+ 2\sum_{k}\biggl[\left(J_1S_2-J_2S_1\right)a^{\dagger}_ka_k + \left(J_1S_1-J_2S_2\right)b^{\dagger}_kb_k \biggr]\\
+ J_2S_1\left(e^{-2ik\delta}a_ka^{\dagger}_k + e^{2ik\delta}a^{\dagger}_ka_k \right) + J_2S_2\left(e^{-2ik\delta}b_kb^{\dagger}_k + e^{2ik\delta}b^{\dagger}_kb_k \right)
\end{multline}

In matrix form it is written as,
\begin{equation*}
\hspace{-1.0cm}
\footnotesize{
\begin{pmatrix}
 J_1S_2 - J_2S_1(1 - \mr{cos}(k) ) & 0 & 0 & J^-\sqrt{S_1S_2}\mr{cos}(k/2) \\
 0 &  J_1S_1 - J_2S_2(1 - \mr{cos}(k) ) & J^-\sqrt{S_1S_2}\mr{cos}(k/2) & 0 \\
 0 & J^+\sqrt{S_1S_2}\mr{cos}(k/2) & J_1S_2 - J_2S_1(1 - \mr{cos}(k) )  & 0 \\
 J^+\sqrt{S_1S_2}\mr{cos}(k/2) & 0 & 0 &J_1S_1 - J_2S_2(1 - \mr{cos}(k) )
 \end{pmatrix} }
\end{equation*}
Boguliubov transfromation, followed by diagonalization gives 4 modes,

\begin{equation}
    \omega_{1k} = \omega_{4k}= \frac{1}{2}\left\{J\left(S_1 - S_2\right)  J_2\left(-S_1 + S_2\right)\left(1-\mr{cos}(k) \right) - \Omega_k\right\}
\end{equation}
and 
\begin{equation}
    \omega_{2k} = \omega_{3k} = \frac{1}{2}\left\{J\left(S_1 - S_2\right) + J_2\left(-S_1 + S_2\right)\left(1-\mr{cos}(k) \right) + \Omega_k\right\}
\end{equation}
where

\subsection{Sublattice Magnetization}
The magnetization for the $A$ sublattice can be calculated as(Ref)

\begin{equation}
    M_a = S_1 - \langle {a^{\dagger}}_{i=1}a_{i=1}\rangle 
    = S_1 - \frac{1}{N_{uc}}\sum_{k,k'}\langle a^{\dagger}_{k}a_{k'}\rangle
\end{equation}
Where $N_{uc}$ is the number of atoms in the unit cell
Using the Bogoliubov transformation defined before $\bm{A}_k = \bm{V}_k .\bm{\Tilde{A}}_k$, we get
\begin{equation}
    M_a
    = S_1 - \frac{1}{N_{uc}}\sum_{k,k',m,n} V_k(1,n)^* V_k(1,m)
    \Tilde{A}^{\dagger}_k(n)\Tilde{A}_k(m)
\end{equation}
When acting on the ground state, only the terms of the form $\alpha_k{\alpha^{\dagger}}_{k}$ is the Bogoliubov basis will have non-zero contribution.
So it is sufficient to sum over only terms $m=n=3,4$ and $k=k'$.
Hence,
\begin{align}{\label{eq:mag_sw}}
    M_a
    &= S_1 - \frac{1}{N_{uc}}\sum_{k,k'}\sum_{m=3,4} V_k(1,n)^* V_k(1,n)
    \Tilde{A}^{\dagger}_k(n)\Tilde{A}_k(n) \notag\\
    &= \left(S_1 + \frac{1}{2}\right) \notag\\
    &+\frac{1}{2}\bigint_{-\pi}^{\pi}{dk\frac{\left(J(S_1+S_2) + \sqrt{-4\left(J^2 + D^2\right)S_1S_2
    \mathrm{cos}^2(k/2) + J^2(S_1+S_2)^2}\right)^2}{4(J^2+D^2)S_1S_2
    \mathrm{cos}^2(k/2)}}
\end{align}
Similarly,
\begin{align}
    M_b
    &= \frac{1}{N_{uc}}\sum_{k,k'}\sum_{m=3,4} V_k(2,n)^* V_k(2,n)
    \Tilde{A}^{\dagger}_k(n)\Tilde{A}_k(n) - S_2 - \frac{1}{2}
\end{align}
The first term of the RHS is equal to the negative of the second term in the RHs of Eqn \ref{eq:mag_sw}